\documentclass[aps,prx,twocolumn,floatfix,footinbib,notitlepage,groupaddress,showpacs]{revtex4-1}
\usepackage{times,color,amsfonts,amssymb,amsbsy,amsthm,amsmath,graphics,graphicx,hyperref,bm,bbm,makecell,mathtools,dsfont,upgreek,fancyhdr,multirow,newtxmath,subfigure}

\DeclareFontFamily{OT1}{pzc}{}
\DeclareFontShape{OT1}{pzc}{m}{it}%
{<-> s * [1.25] pzcmi7t}{}
\DeclareMathAlphabet{\mathpzc}{OT1}{pzc}%
{m}{it}
\hypersetup{colorlinks,linkcolor={blue},citecolor={blue},urlcolor={blue}}
\makeindex

\let\oldsqrt\sqrt
\def\sqrt{\mathpalette\DHLhksqrt}
\def\DHLhksqrt#1#2{%
\setbox0=\hbox{$#1\oldsqrt{#2\,}$}\dimen0=\ht0
\advance\dimen0-0.2\ht0
\setbox2=\hbox{\vrule height\ht0 depth -\dimen0}%
{\box0\lower0.4pt\box2}}

\begin{document}

\title{Capacity-approaching quantum repeaters for quantum communications}
\author{Masoud Ghalaii}
\affiliation{Department of Computer Science, University of York, York YO10 5GH, United Kingdom}
\author{Stefano Pirandola}
\affiliation{Department of Computer Science, University of York, York YO10 5GH, United Kingdom}

\begin{abstract}
In present-day quantum communications, one of the main problems is the lack of a quantum repeater design that can simultaneously secure high rates and long distances. Recent literature has established the end-to-end capacities that are achievable by the most general protocols for quantum and private communication within a quantum network, encompassing the case of a quantum repeater chain. However, whether or not a physical design exists to approach such capacities remains a challenging objective. Driven by this motivation, in this work, we put forward a design for continuous-variable quantum repeaters and show that it can actually achieve the feat. We also show that even in a noisy regime our rates surpass the Pirandola-Laurenza-Ottaviani-Banchi (PLOB) bound. Our repeater setup is developed upon using noiseless linear amplifiers, quantum memories, and continuous-variable Bell measurements. We, furthermore, propose a non-ideal model for continuous-variable quantum memories that we make use of in our design. We then show that potential quantum communications rates would deviate from the theoretical capacities, as one would expect, if the quantum link is too noisy and/or low-quality quantum memories and amplifiers are employed. 
\end{abstract}

\maketitle

\section{INTRODUCTION}

One of the most developed applications of quantum technologies is assuredly quantum communications \cite{Kimble:QuInternet,Pirandola:Nat_QInternet,Razavi:Book2018,Pirandola:Review15}, within which the most advanced primitive is quantum key distribution (QKD)---the artistry of sharing secret key strings to two, or more, distant stations in the present of untrusted adversaries \cite{Gisin:RevQC2002,Pirandola:RevQKD2019}.
The security of QKD depends on the kinds of assumption one puts on the adversaries, which in turn define the quantum communication link between trusted stations, as well as adversaries' power class (namely, individual, collective, and coherent). For instance, conveniently, a quantum communication link is described by a bosonic thermal-loss channel. 
Furthermore, these assumptions impose limits on the maximum transmission distance over which legitimate parties can securely generate a key.

On the other hand, in order to establish a global quantum communications network \cite{Razavi:Book2018}, i.e., quantum internet \cite{Kimble:QuInternet, Azuma:QuInternet,Pirandola:Nat_QInternet}, enabling QKD protocols (resistant to loss and noise) to be executed over distances beyond one thousand kilometres is of great practical importance. 
To go that far, naturally stemming from information theory \cite{Cover,Gamal}, one can build a quantum repeater \cite{Briegel:DVQRs1998,Munro:QRreview2015} by breaking down a long-distance communication link into two, or more, short links, and  linking them by means of proper joint measurements. 
While there are quantum repeater proposals  for quantum communication networks based on discrete variable systems \cite{Holevo,Nielsen_Chuang,Watrous}, which rout back to the seminal work of Briegel \emph{et al.} \cite{Briegel:DVQRs1998}, we focus on quantum repeaters for continuous variable (CV) systems \cite{Weedbrook:GaussQI2012,Braunstein:QICVRev2005} that have recently been brought into attention \cite{Dias:CVQR2017,Dias:CVQR2020,Furrer:CVQR2018,Dias:ComparingQRs2019,Seshadreesan:QSCVQR2020,Pirandola:EndtoEnd2019}. 
We note that one way to increase the reach of a CV-QKD protocol \cite{Grosshans:GG02PRL,Grosshans:GG02Nature} is to use quantum noiseless linear amplifiers (NLAs) \cite{Blandino:idealNLA2012}. However, they can only improve the secure distance for few tens of kilometres \cite{Blandino:idealNLA2012,Ghalaii:JSTQE2020,Ghalaii:JSAC2020}. Hence, one practicable idea to explore is to  concatenate such NLA-improved links.  

To distribute entanglement to a farther distance, compared to what we could have achieved otherwise, is the ultimate goal of a quantum repeater. 
More precisely, the goal is to beat the PLOB bound~\cite{Pirandola_PLOB17}, which cannot be exceeded even by the so-called adaptive local operations and classical communication strategies \cite{Pirandola:AdLOCCs}.
Accomplishing a goal as such requires three indispensable components: entanglement distribution, entanglement distillation or purification, and entanglement swapping. 
In CV systems, a natural entangled state to use---at the entanglement distribution stage---is a two-mode squeezed vacuum (TMSV) state ~\cite{Weedbrook:GaussQI2012}, as it has been exploited in almost all previous studies on CV quantum repeaters \cite{Dias:CVQR2017,Dias:CVQR2020,Furrer:CVQR2018,Seshadreesan:QSCVQR2020,Dias:ComparingQRs2019}.  
For the entanglement purification stage, one can make use of NLAs, such as quantum scissors \cite{Seshadreesan:QSCVQR2020}, or non-Gaussian entanglement distillation protocols, e.g., photon-added schemes \cite{Furrer:CVQR2018}.  
Finally, similar to discrete-variable quantum repeaters \cite{Briegel:DVQRs1998,Bruschi:DVQR2014}, where a Bell state measurement swaps entanglement between two neighbouring links, one needs to apply a Bell-like measurement in order to jointly measure two nodes in a CV quantum repeater. 
This task, for instance, can be performed via a CV quantum relay \cite{Braunstein_CVtele,Spedalieri:BellLike2013}, or a more elaborate apparatus that needs single photon injection \cite{Furrer:CVQR2018}.

There nonetheless seems to be a fundamental lack in designing a quantum repeater: either the Bell(-like) measurement or the purification stage, if not both, is probabilistic. 
This seemingly unavoidable impediment impels us to equip quantum repeater nodes with ancillary components, e.g., quantum memories \cite{Lvovsky:OptQMs2009,Simon:QMReview2010}, whose presence have been proven efficacious, if not essential \cite{Munro:QCwithoutQR2012,Azuma:All-photonicQR2015}, to improve performance of quantum repeaters. 
To our knowledge, all previous studies on CV quantum repeaters \cite{Dias:CVQR2017,Dias:CVQR2020,Furrer:CVQR2018,Dias:ComparingQRs2019,Seshadreesan:QSCVQR2020} have considered quantum memories as a key component, even-though they were assumed ideal.

In the quest for the ultimate performance of a quantum communication network, including a quantum repeater chain, one of us~\cite{Pirandola:EndtoEnd2019} has established the end-to-end capacities achievable by the most powerful protocols under general routing strategies. In Ref.~\cite{Pirandola:EndtoEnd2019}, upper bounds are derived by extending the techniques first devised from point-to-point to repeater-assisted quantum communications \cite{Pirandola_PLOB17}.
On the other hand, for the lower bounds it uses sessions of QKD over each link followed by key composition via one time pad. Also, for entanglement distribution it assumes optimal entanglement distillation over each link, and then entanglement swapping.
There, for instance, it was shown that the capacity of a quantum repeater chain  with the most important optical communications link as its basic link, i.e., the bosonic lossy channel, is given by $-\log_2(1-\eta^{1/N})$, bits per channel use, where $\eta$ is the channel loss and $N$ the number of links in the chain. 
Since Ref.~\cite{Pirandola:EndtoEnd2019} aimed at establishing capacities and sought for benchmarks, it allowed for the most general quantum operations and measurements, regardless of their practicality. 
Hence, whether or not one can design a practical setup that can beat, or even come close to, such bounds is an exigent query to study.

\begin{figure}[b]
	\centering
	\includegraphics[scale=0.85]{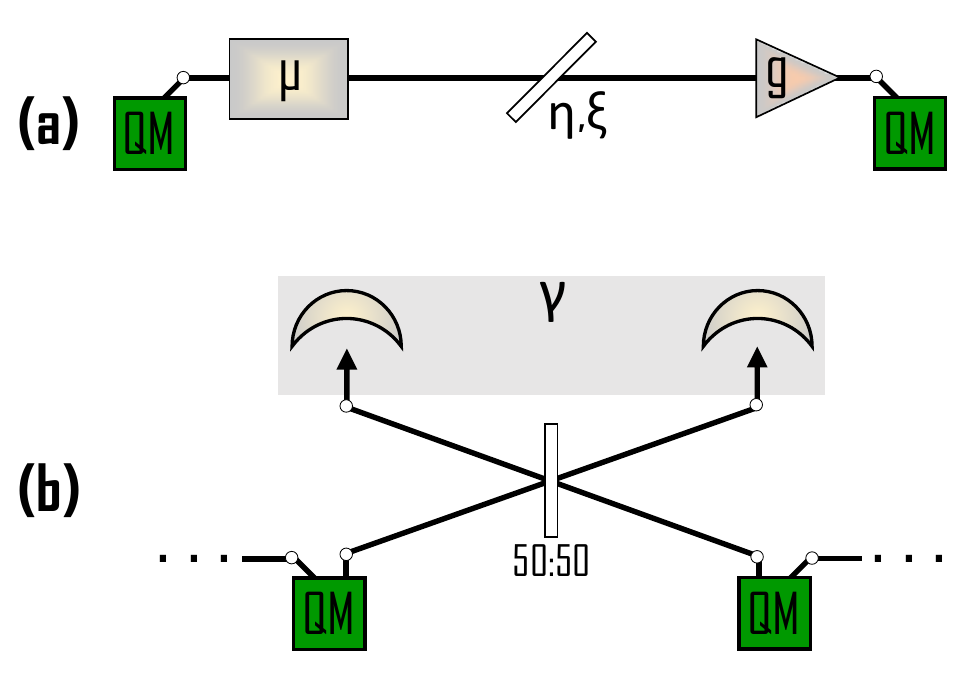}
	\caption{\textbf{(a) Basic module.} A TMSV source generates a bipartite entangled state, with variance $\mu$, of which one light mode is stored in a quantum memory whereas the other is left propagating through a thermal-loss quantum link described by the pair $(\eta,\xi)$. At the end of the link, upon a successful NLA operation, with amplification gain $g$, the amplified state is stored in another quantum memory.
	\textbf{(b) Continuous variable quantum relay.} Adjacent quantum memories get connected via a CV Bell detection, which includes a balanced beam splitter followed by two conjugate homodyne detectors. The outcome of the measurement, $\mathbf{\gamma}$, is then kept for later use. Green QM boxes indicate quantum memories. }
	\label{fig:basiclink_relay}
\end{figure}

In order to study the ultimate practical performance of a quantum repeater, we notice that a repeater chain generally has two main compartments.
Indeed, one is the basic quantum module linking each two adjacent repeater stations. 
A realistic communication link of this kind is a thermal-loss channel that has a TMSV source at its one (transmitter) end. 
Nevertheless, such a link can be  backed up with a quantum amplifier at its other (receiver) end, or quantum memories at both ends. 
We note that while the use of ideal and non-ideal NLAs in quantum communications with CV systems have rather extensively been studied, both in single-link and repeater scenarios \cite{Blandino:idealNLA2012,Dias:CVQR2017,Dias:CVQR2020,Seshadreesan:QSCVQR2020,Ghalaii:JSTQE2020}, quantum memories, which are as crucial, have been overlooked. 
In general thus the basic module, which we indicate by $\mathpzc{E}=\mathpzc{E}(\mu,\eta,\xi,g)$, can be described as shown in Fig.~\ref{fig:basiclink_relay}(a).
The other key compartment is a joint connecting measurement between two neighbouring links, or more precisely neighbouring quantum memories, that is preformed at each station of a repeater.
While in discrete-variable scenarios one would typically perform a Bell state measurement \cite{Bruschi:DVQR2014}, in continuous variable protocols we would apply a CV version of it, shown in Fig.~\ref{fig:basiclink_relay}(b), which includes a balanced beam splitter followed by two (conjugate) homodyne detection modules~\cite{Braunstein_CVtele,Spedalieri:BellLike2013}.

In this paper, we study the performance of continuous-variable quantum repeaters that run under the following assumptions. We assume that the repeater is made of a concatenation of  thermal-loss links, shown schematically in Fig.~\ref{fig:basiclink_relay}(a). We assume {\em non-ideal} CV quantum memories, i.e., quantum memories whose coherence times are finite and comparable to operational times \cite{Op_times} of the system.
To do so, we  model  quantum memories for CV systems based on decoherence of quantum states in a thermal bath \cite{Adesso:QMdecoh2006}. 
We also allow for assisting the basic links with NLAs. Indeed, we can choose not to use NLAs, especially at short distances, as it is often the case \cite{Blandino:idealNLA2012,Ghalaii:JSTQE2020,Ghalaii:JSAC2020}. In this study, we take ideal NLAs as were formerly used in Refs.~\cite{Blandino:idealNLA2012,Dias:CVQR2017}. Finally, as a realistic joint connecting measurement, we take a CV Bell detection as shown in Fig.~\ref{fig:basiclink_relay}(b).

Under the above assumptions, we find a recursive equation for the covariance matrix of the end-to-end state of the repeater chain with an arbitrary number of links. Based on the obtained covariance matrix, we then provide general bounds for secret key rates at different repeater depths. 
Remarkably, we show that obtainable rates by our design can in principle approach the repeater capacities derived in Ref.~\cite{Pirandola:EndtoEnd2019}. 
Moreover, our study quantitatively shows how non-ideal quantum memories effect the obtained key rates. Our results certify that possessing quantum memories with a certain operational coherence time is a \emph{sine qua non} to preserve the prospect of having efficient CV quantum repeaters.

\section{THE QUANTUM REPEATER PROTOCOL}

We assume that the far-end parties, Alice and Bob, are connected via a repeater chain of depth $n$, which involves $N=2^n$ identical basic quantum transmission links; see Fig.~\ref{fig:QR}. 
Such a chain then has a total length of $L=2^nL_0$, where $L_0$ is the point-to-point distance of each basic link. 
While one can describe each basic link by a thermal-loss link with transmittance $\eta$ and excess noise $\xi$, we, more generally, take a quantum memory and NLA-assisted thermal-loss link, as shown in Fig.~\ref{fig:basiclink_relay}(a). 

If we prefer to use NLAs, which are non-deterministic operations, we would inevitably make the whole repeater setup probabilistic. This is where high-quality quantum memories can be useful, where heralded successful quantum links are stored in a couple of quantum memories until their adjacent links are also announced successful. 
One can also benefit from quantum memories if the Bell state measurements are non-deterministic, as studied in the context of discrete-variable quantum repeaters \cite{Bruschi:DVQR2014}. 
In this study, however, we use CV Bell detection modules, see Fig.~\ref{fig:basiclink_relay}(b), which are deterministic.

In the following, we first study the detail of the CV quantum repeater with ideal quantum memories. Next, we introduce our model for quantum memories and, subsequently, explore the performance of the repeater when such memories are in operation.  

\subsection{Properties of a basic link} 
Conveniently, one would only need the covariance matrix (CM) of the two ends of a basic link to fully solve a Gaussian system, e.g., that given in Fig.~\ref{fig:basiclink_relay}(a), wherein all components are (assumed) Gaussian. In particular, here we assume that the NLA, ideally, does a Gaussian, yet non-deterministic, operation. 
In fact, it does $|\alpha\rangle \rightarrow |g\alpha\rangle$, where $|\alpha\rangle$ is a coherent state, with some probability \cite{Pandey_Qlimits_Amp2013,Blandino:idealNLA2012}. In the following, we will make use of the Gaussianity of the system.

\begin{figure}[t]
	\includegraphics[scale=0.62]{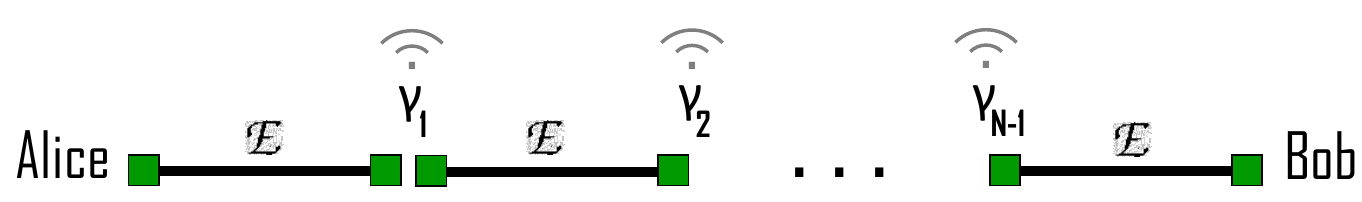}
	\caption{\textbf{Quantum repeater chain.} The distance between far-end parties, Alice and Bob, is divided into $N$ basic links, each is backed up with a pair of quantum memories (green squares) located at each station, where 
	by applying a CV Bell detection on each pair, unconnected links get connected. For proper adjustments, the outcome of each Bell detection, $\gamma_j$, is then aired to other stations via classical communications.}
	\label{fig:QR}
\end{figure}

Two-mode squeezed vacuum state is perhaps the quintessential two-mode state that has extensively been used in CV quantum information \cite{Weedbrook:GaussQI2012,Braunstein:QICVRev2005}. The CM of such a state, generated by the $\mu$ box in Fig.~\ref{fig:basiclink_relay}(a), has the form 
\begin{align}
\label{CM:TMSV}
\mathbf{V}=
\left(\begin{array}{cc}
a \mathbf{I} & c \mathbf{Z} \\
c \mathbf{Z} &  b \mathbf{I}
\end{array}\right),
\end{align}
where $\mathbf{I}:=\text{diag}(1,1)$ and $\mathbf{Z}:=\text{diag}(1,-1)$ are Pauli matrices,  $a=b=\mu$ is the quadrature variance at both modes, and $c=\sqrt{\mu^2-1}$ is correlation/covariance between them.
Next, we assume that one mode of this TMSV state propagates thorough a thermal-loss channel, characterized by the channel transmittance $\eta$ and excess noise $\xi$. 
Upon a successful NLA operation, the CM of the state at the two ends of a link that would be stored in the quantum memories is given by \cite{Blandino:idealNLA2012} 
\begin{align}
\label{CM:basiclink}
\mathbf{V}_0=
\left(\begin{array}{cc}
\mathbf{a}_0 \mathbf{I} & \mathbf{c}_0 \mathbf{Z} \\
\mathbf{c}_0 \mathbf{Z} &  \mathbf{b}_0 \mathbf{I}
\end{array}\right),
\end{align}
where the triplet $(\mathbf{a}_0,\mathbf{b}_0,\mathbf{c}_0)$ is defined in App.~\ref{app:triplet}. 
{We hence describe the basic link via $\mathpzc{E}=\mathpzc{E}(\mu_g,\eta_g,\xi_g,1)$, given in Eq.~\eqref{equivalent_eqs}}, whose CM is given by Eq.~\eqref{CM:basiclink}.  
Therefore, the CV repeater chain of depth $n$ between Alice and Bob is constructed upon linking $2^n$ $\mathpzc{E}$'s by executing CV Bell measurements, as shown in Fig.~\ref{fig:QR}. 
{The equivalence relation $\mathpzc{E}(\mu,\eta,\xi,g)=\mathpzc{E}(\mu_g,\eta_g,\xi_g,1)$ should be taken with caution, for the reason that NLAs can result in unreliable outcomes. More specifically, as discussed in Ref.~\cite{Blandino:idealNLA2012}, a set of quantum channel parameters and amplification gain $\lambda_g=\sqrt{\frac{\mu_g-1}{\mu_g+1}}$, which describes the equivalent TMSV state, can be greater than 1. We back to this point in Sec.~\ref{sec:results}. }

\subsection{Recursive equation for an arbitrary number of basic links}
Assume Alice and Bob have all $N$ basic links, $\mathpzc{E}(\mu,\eta,\xi,g)$'s, of the repeater chain ready to be executed. That is, succeeding successful NLA events at all stations, each state is stored in a couple of perfect, i.e., lossless and noise free, quantum memories.
The next stage is then to perform joint Bell-like measurements. 
We apply the CV Bell measurement in Fig.~\ref{fig:basiclink_relay}(b) to each two neighbouring quantum memories. This leads to obtaining a recursive equation for the end-to-end CM of the repeater setup (see App.~\ref{app:CVRelay} for details)
\begin{align}
\label{CM:recursive}
\mathbf{V}_n=
\left(\begin{array}{cc}
\mathbf{a}_n \mathbf{I} & \mathbf{c}_n \mathbf{Z} \\
\mathbf{c}_n \mathbf{Z} &  \mathbf{b}_n \mathbf{I}
\end{array}\right),
\end{align}
where
\begin{align}
\begin{cases}
\mathbf{a}_n=\mathbf{a}_{n-1} - \frac{\mathbf{c}_{n-1}^2 }{\mathbf{a}_{n-1} + \mathbf{b}_{n-1} }  \\
\mathbf{b}_n= \mathbf{b}_{n-1} - \frac{\mathbf{c}_{n-1}^2 }{\mathbf{a}_{n-1} + \mathbf{b}_{n-1} }   \\
\mathbf{c}_n= \frac{\mathbf{c}_{n-1}^2 }{\mathbf{a}_{n-1} + \mathbf{b}_{n-1} } .
\end{cases}
\end{align}

We see that the far-end CM of the repeater at depth $n$ depends on only that at depth $n-1$. Luckily, without great effort, this can numerically be managed.

\subsection{Continuous-variable quantum memories}
Quantum memories are supposed to keep quantum states intact over a fairly good amount of time.
A realistic quantum memory, nevertheless, can fail to perpetuate the quantum features of the stored states.
This occurs in a process called decoherence, or decay, which may not necessarily be Gaussian. Following Ref.~\cite{Adesso:QMdecoh2006}, we model CV quantum memories as devices that change the CM of Gaussian states as time elapses as 
\begin{align}
\mathbf{V}(t)=\mathbf{\Gamma}(t) \mathbf{V}(0)+ (\mathbf{I} \oplus \mathbf{I}-\mathbf{\Gamma}(t)) \mathbf{V}(\infty),
\end{align}
where $\mathbf{\Gamma}(t)=\oplus_j e^{-t/\tau_{\rm c} } \mathbf{I}$, with $\tau_{\rm c} $ being the coherence time of the memory. 
In addition, $\mathbf{V}(0)$ is the initial CM, at time $t=0$, whereas $\mathbf{V}(\infty)$ is a diagonal matrix and, ideally, is proportional to identity matrix. 
A realistic quantum memory would possibly add noise to the quantum system \cite{Jensen:CVQMs2011}. 
Such a noise can be simulated by the following modification 
\begin{align}
\mathbf{V}(\infty)=\oplus_j (1+\xi_{\rm QM}) \mathbf{I},
\end{align}
where $\xi_{\rm QM}$ quantifies the amount of noise that a quantum memory adds to the stored CV states. 
{Note that we do not assume an inner structure for the QMs and simply model a generic QM for continuous-variable states by considering it---based on Ref.~\cite{Adesso:QMdecoh2006}---as a thermal-loss channel. However, while Ref.~\cite{Adesso:QMdecoh2006} let the stored states, at long enough time, to become vacuum states, we consider an even worse case by adding excess noise, where the stored states become thermalized states with a variance greater than 1.  
}

Therefore, we realize that a CV quantum memory functions as a thermal-loss channel that is characterized by means of $\mathbf{\Gamma}(t)$ and $\mathbf{V}(\infty)$, which correspond, respectively, to loss and excess noise.  
We remark that a quantum memory for CV systems have been experimentally demonstrated and tested  \cite{Jensen:CVQMs2011}. Such a quantum memory was shown to keep entanglement of a 6.0~dB TMSV state for 1ms \cite{TMSVdB}. However, due to loss and noise caused by the memory, a fidelity of only 0.52 could have been achieved. 

As a result of our model, the triplet $(\mathbf{A}_0(0),\mathbf{B}_0(0),\mathbf{C}_0(0))$ that defines an initial CM at time $t=0$ would change to the following triplet 
\begin{align}
\label{timedep_CMelements}
\begin{cases}
\mathbf{A}_0(t)=1+\xi_{\rm QM}+(\mathbf{A}_0(0)-1-\xi_{\rm QM})e^{-t/\tau_{\rm c}} \\
\mathbf{B}_0(t)=1+\xi_{\rm QM}+(\mathbf{B}_0(0)-1-\xi_{\rm QM})e^{-t/\tau_{\rm c}} \\
\mathbf{C}_0(t)=\mathbf{C}_0(0)e^{-t/\tau_{\rm c}},
\end{cases}
\end{align}
at time $t$.
Asymptotically in time, i.e., $t \rightarrow \infty$, as one would expect, we get $\mathbf{A}_0(t)=\mathbf{B}_0(t)=1+\xi_{\rm QM}$ and $\mathbf{C}_0(t)=0$, where we are left with a noisy thermalized bipartite state and zero quantum correlation between quadratures.
We later use this model to compute achievable secret key rates of CV quantum communication protocols run over a quantum repeater assisted with such non-ideal quantum memories.

\subsection{Recursive equation with quantum memories}
Assume that a two-mode state with a CM given by  Eq.~\eqref{CM:basiclink} is stored, at time $t=0$, in a pair of quantum memories described above. 
As time elapses, the CM varies such that at time $t$ it is given by
\begin{align}
\label{CM:basiclinkQM}
\mathbf{V}_0(t)=
\left(\begin{array}{cc}
\mathbf{a}_0(t) \mathbf{I} & \mathbf{c}_0(t) \mathbf{Z} \\
\mathbf{c}_0(t) \mathbf{Z} &  \mathbf{b}_0(t) \mathbf{I}
\end{array}\right),
\end{align}
where time-dependent elements of the CM are given by Eq.~\eqref{timedep_CMelements}. 

Now before taking on a recursive equation, let us introduce a time notion, $t_{\rm h}$, as a characteristic of a basic link. Assume that $t_{\rm h}$ is the average time that a basic link, i.e., its NLA, is heralded successful.
One can find that, for a basic link of length $L_0$ assisted with an NLA with success probability $P_{\rm succ}$, we have $t_{\rm h}=2L_0/(cP_{\rm succ})$, where $c$ is the speed of light. 
We also let all basic links run in parallel, by which we mean that each link of the repeater is running independently and simultaneously. The objective here is to prepare all links of the repeater chain in a certain state. 
In fact, this is quantum memories that make this scenario doable, since they allow for storing each successful link. 

{Let us also assume that after a unsuccessful attempt NLAs are immediately ready to operate, so that we ignore the time that may take to ``re-load" an NLA.
However, we note that this can be partly, if not fully, covered by the time $t_{\rm h}$. In fact, upon an unsuccessful NLA, the QM can be re-loaded at the time interval when the sender is informed plus the time it takes a new signal reaches the NLA.}

Next, after establishing successful NLA operations over basic links, one can potentially execute all CV Bell measurements simultaneously at all stations. 
Indeed, we can perform a CV Bell measurement on each two adjacent heralded links; however, inevitably, we have to store the resultant state in another pair of quantum memories for some time. In fact, there are various permutations that one can execute or postpone these measurements. In this work, we postpone them all to the moment all links are announced to be successful. 

The worst case one can think of, in a chain with $N$ links, is that $N-1$ links are successful at $t=0$. Notwithstanding, all these successful basic links are required to be stored and await the last one to be heralded successful, which takes $t_{\rm h}$ seconds. Hence, $N-1$ links will suffer memory loss and noise for $t_{\rm h}$ seconds. 
For convenience, so that we can establish a recursive equation, we assume that \emph{all} $N$ basic links go through memory loss and noise for $t_{\rm h}$ seconds. 

Therefore, we end up with $N=2^n$ basic links, each of which is described by $\mathbf{V}_0(t_{\rm h})$. Similar to the case with ideal quantum memories, by executing CV Bell measurements, we obtain the recursive equation for the end-to-end CM of the repeater
\begin{align}
\label{CM:QRQM}
\mathbf{V}_n(t_{\rm h})=
\left(\begin{array}{cc}
\mathbf{a}_n(t_{\rm h}) \mathbf{I} & \mathbf{c}_n(t_{\rm h}) \mathbf{Z} \\
\mathbf{c}_n(t_{\rm h}) \mathbf{Z} &  \mathbf{b}_n(t_{\rm h}) \mathbf{I}
\end{array}\right),
\end{align}
where
\begin{align}
\begin{cases}
\mathbf{a}_n(t_{\rm h})=\mathbf{a}_{n-1}(t_{\rm h}) - \frac{\mathbf{c}_{n-1}^2(t_{\rm h}) }{\mathbf{a}_{n-1} (t_{\rm h})+ \mathbf{b}_{n-1} (t_{\rm h})}  \\
\mathbf{b}_n(t_{\rm h})= \mathbf{b}_{n-1}(t_{\rm h}) - \frac{\mathbf{c}_{n-1}^2(t_{\rm h}) }{\mathbf{a}_{n-1} (t_{\rm h})+ \mathbf{b}_{n-1} (t_{\rm h})}   \\
\mathbf{c}_n(t_{\rm h})= \frac{\mathbf{c}_{n-1}^2(t_{\rm h}) }{\mathbf{a}_{n-1}(t_{\rm h}) + \mathbf{b}_{n-1}(t_{\rm h}) } . 
\end{cases} 
\end{align}

For an arbitrary repeater depth $n$, we can next derive relevant lower or upper bounds on the key rate from the above equation.

 \begin{figure*}[t]
    \centering
    \begin{subfigure}
        \centering
        \includegraphics[scale=0.30]{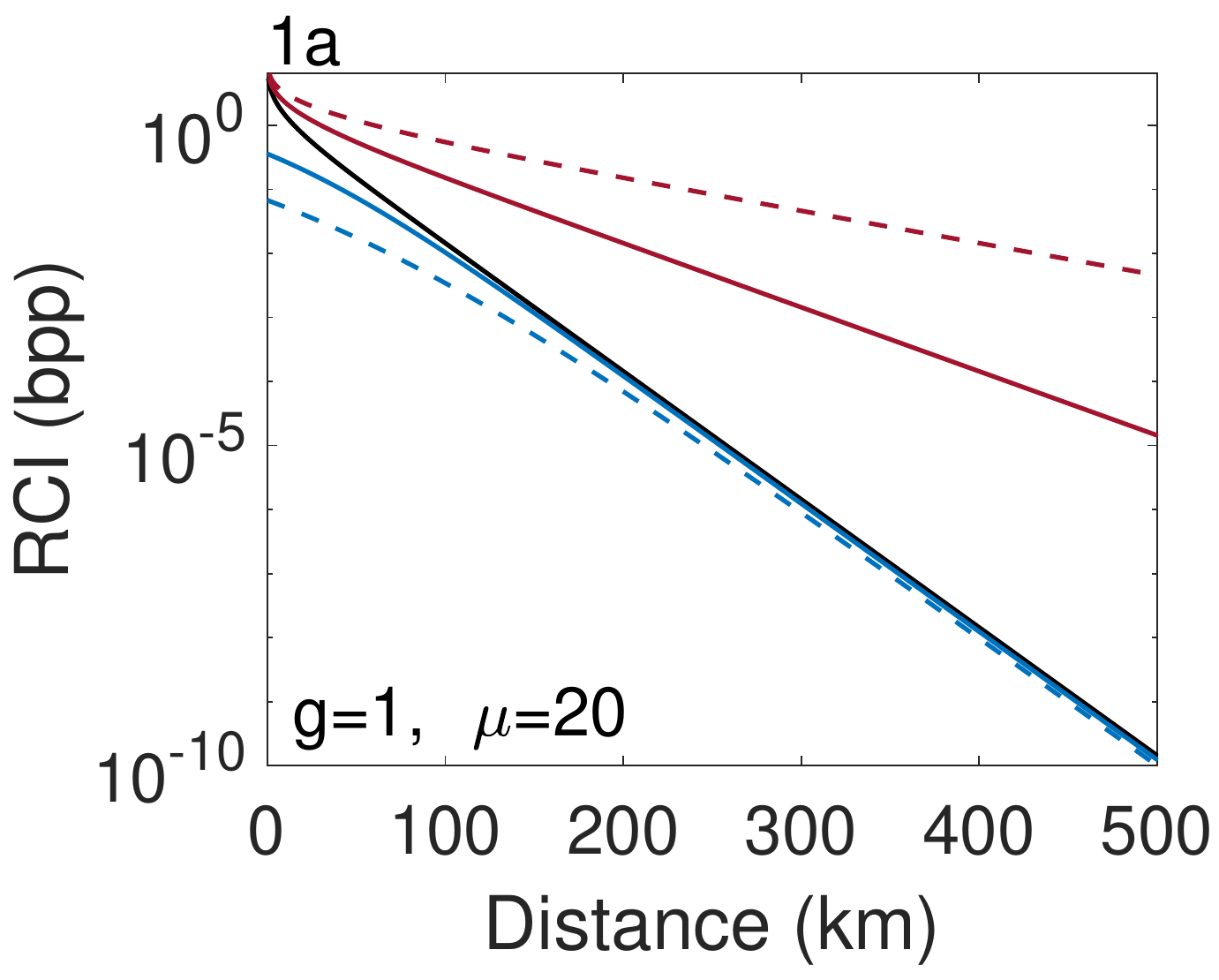} 
    \end{subfigure}
    \begin{subfigure}
        \centering
        \includegraphics[scale=0.30]{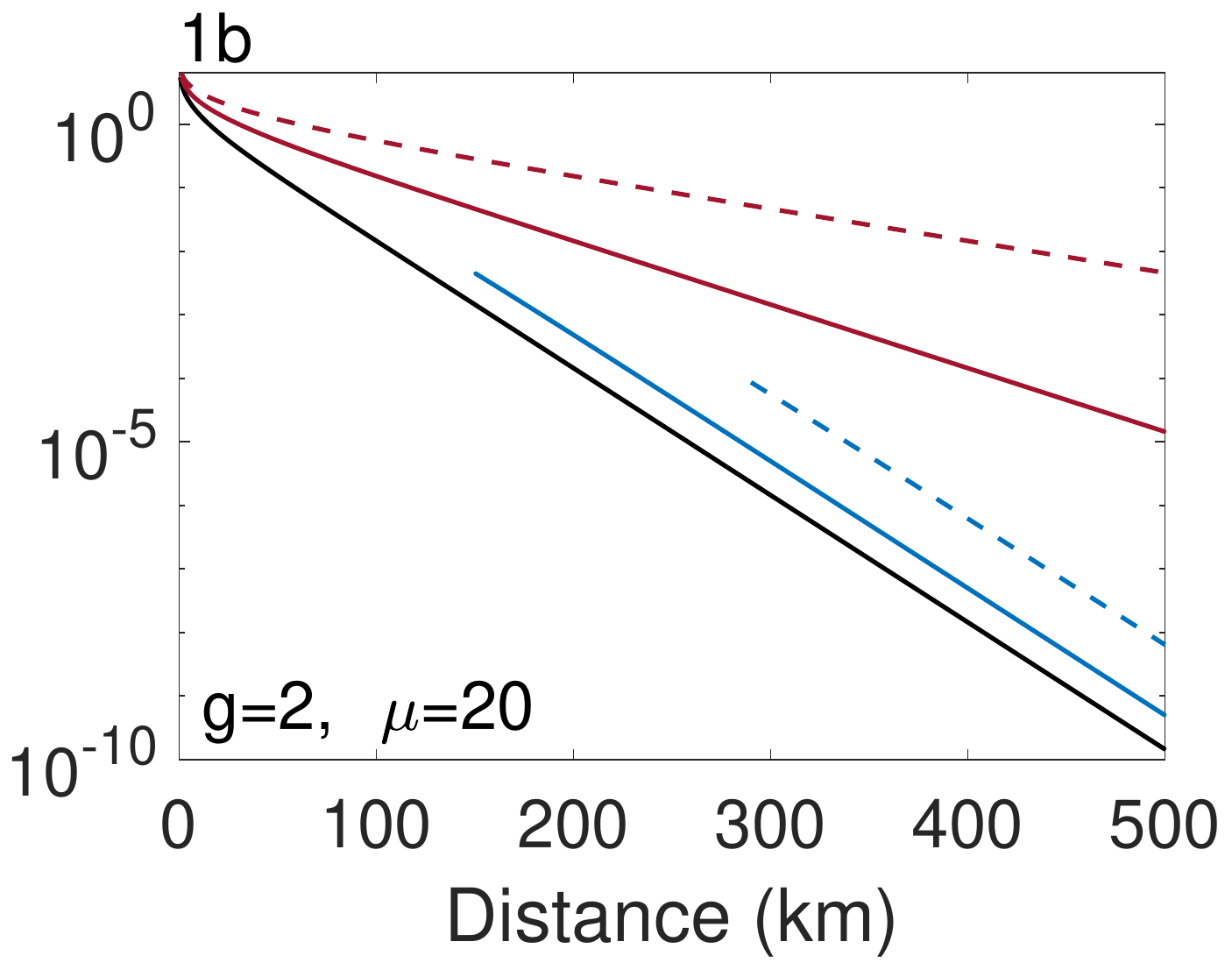} 
    \end{subfigure}
    \begin{subfigure}
    \centering
        \includegraphics[scale=0.30]{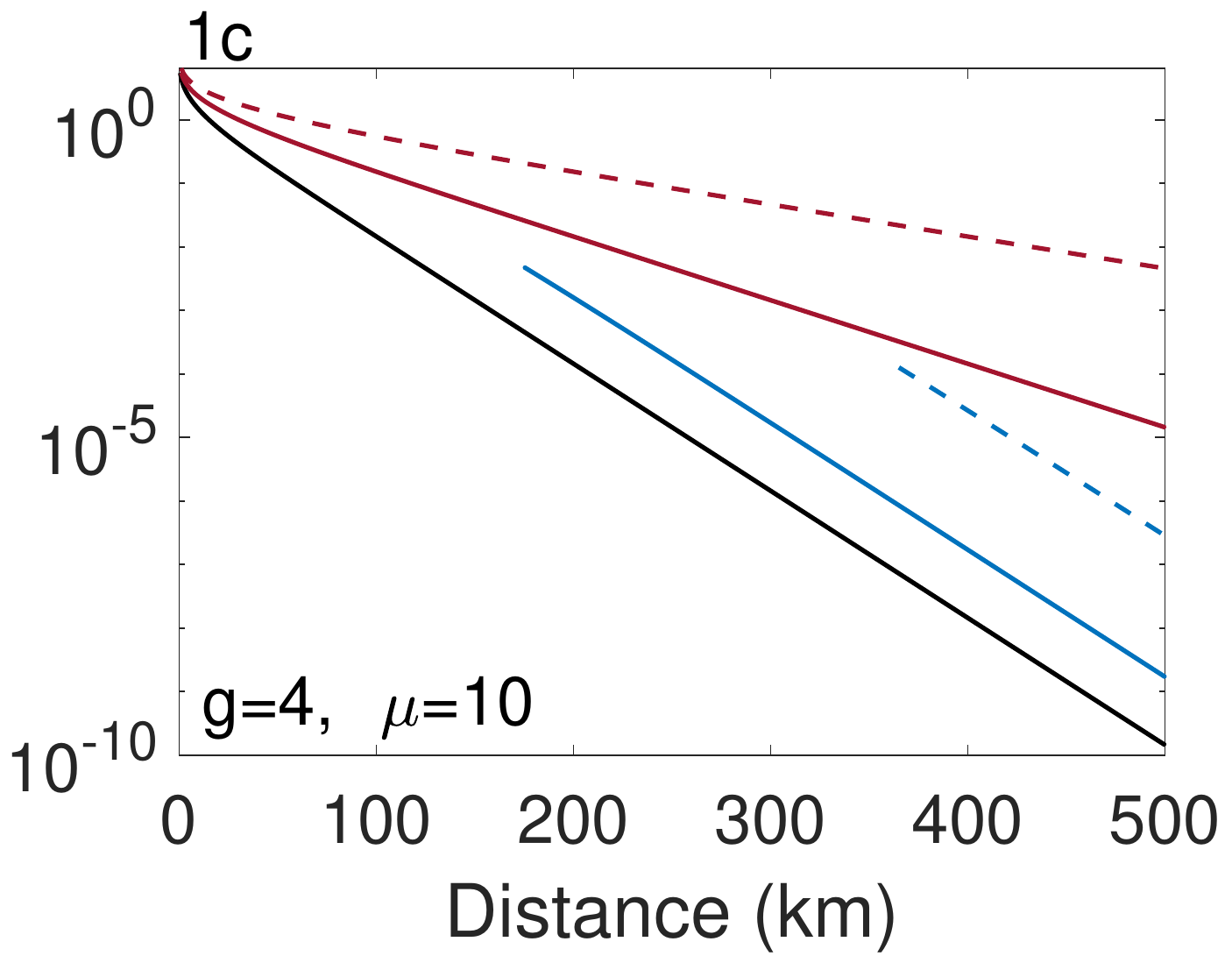} 
    \end{subfigure}
    \begin{subfigure}
    \centering
        \includegraphics[scale=0.30]{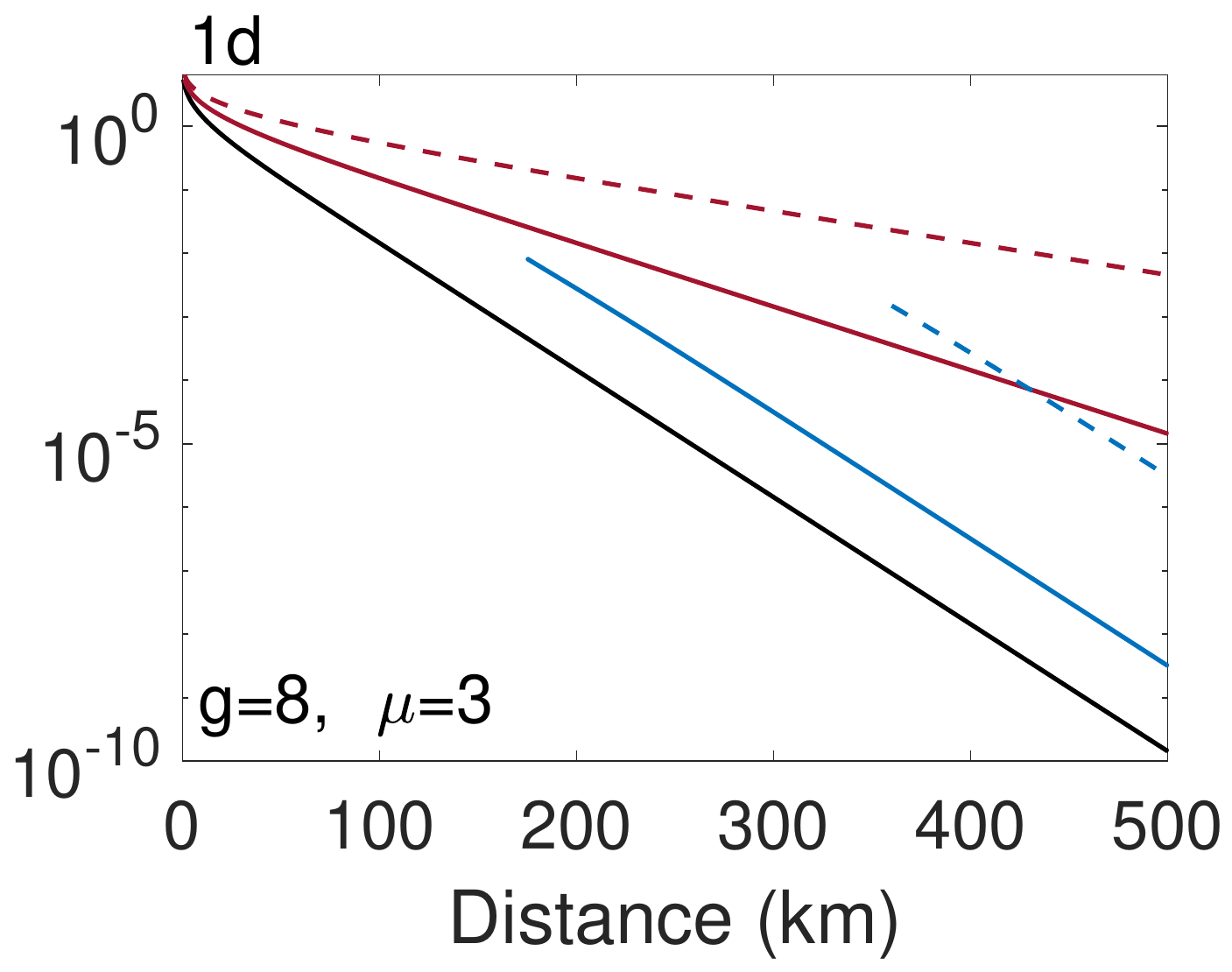} 
    \end{subfigure}
    \begin{subfigure}
        \centering
        \includegraphics[scale=0.30]{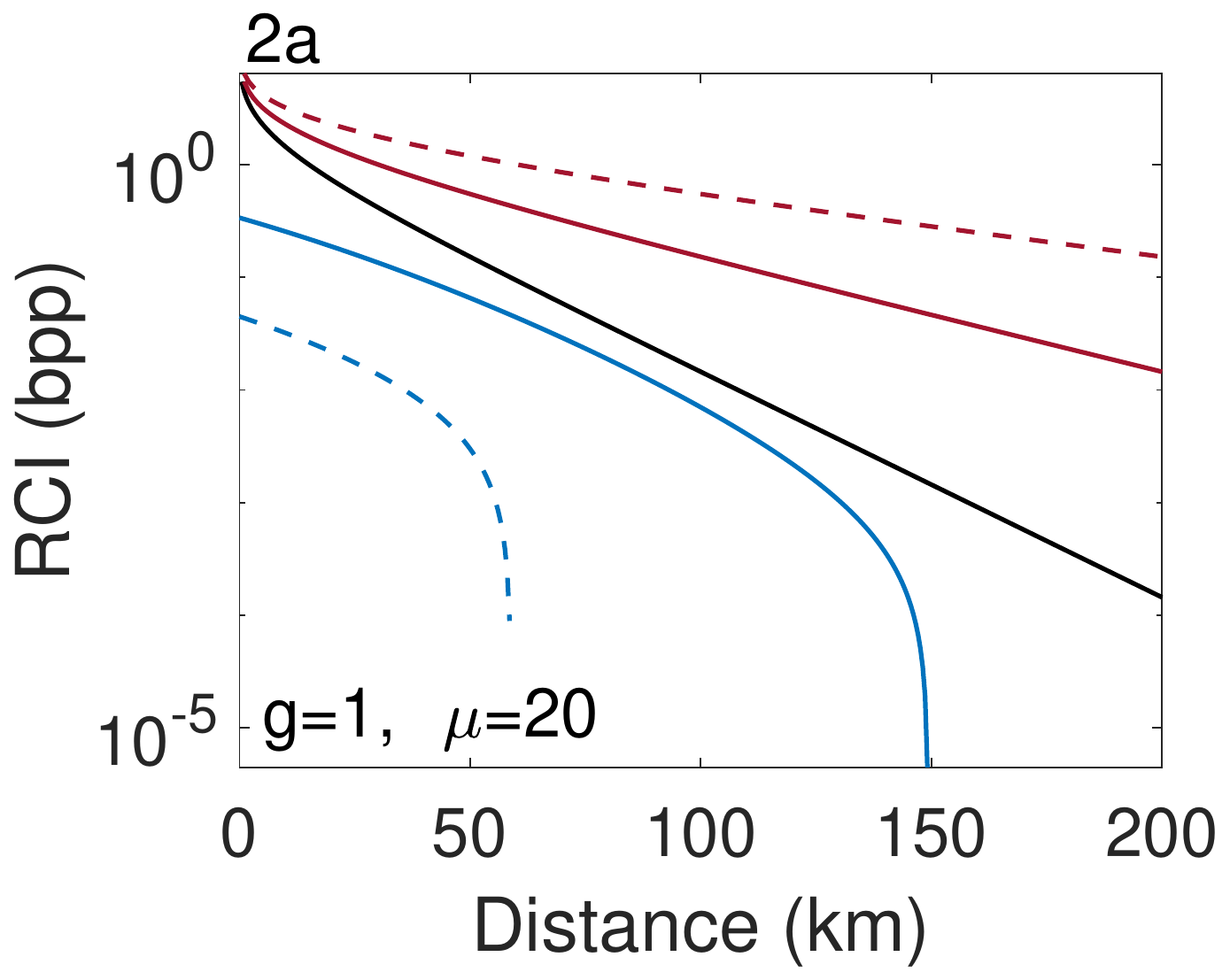} 
    \end{subfigure}
    \begin{subfigure}
        \centering
        \includegraphics[scale=0.30]{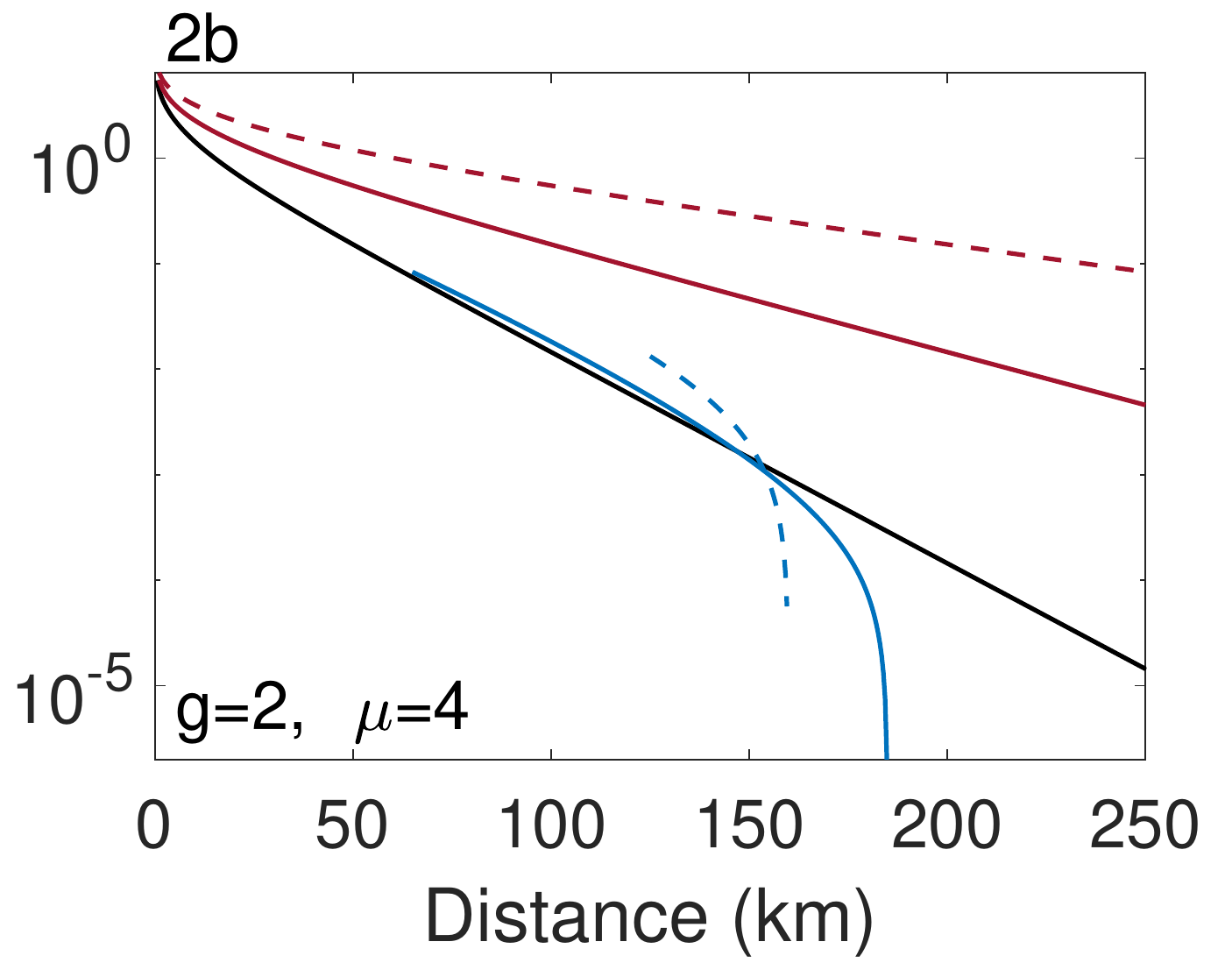} 
    \end{subfigure}
    \begin{subfigure}
    \centering
        \includegraphics[scale=0.30]{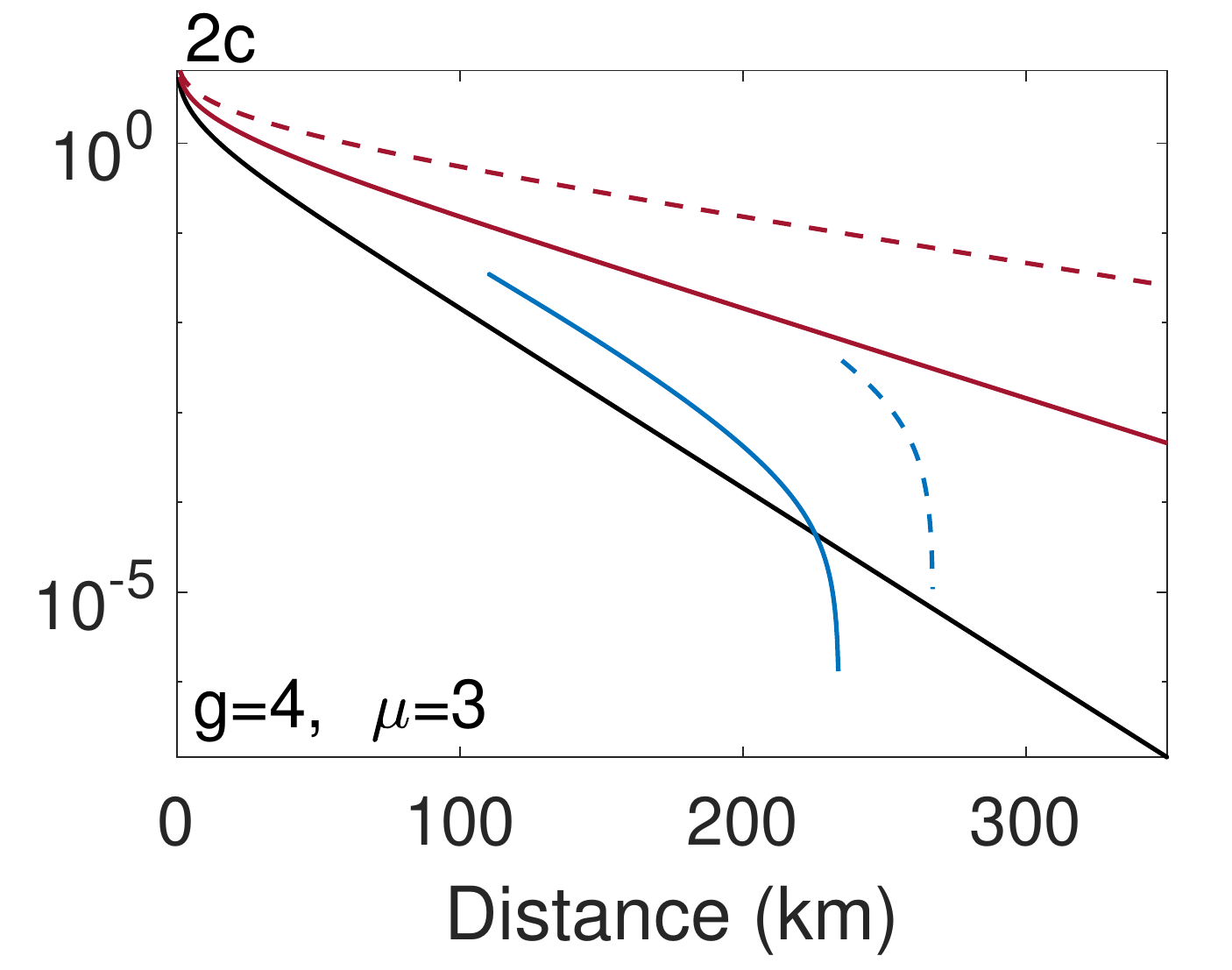} 
    \end{subfigure}
    \begin{subfigure}
    \centering
        \includegraphics[scale=0.30]{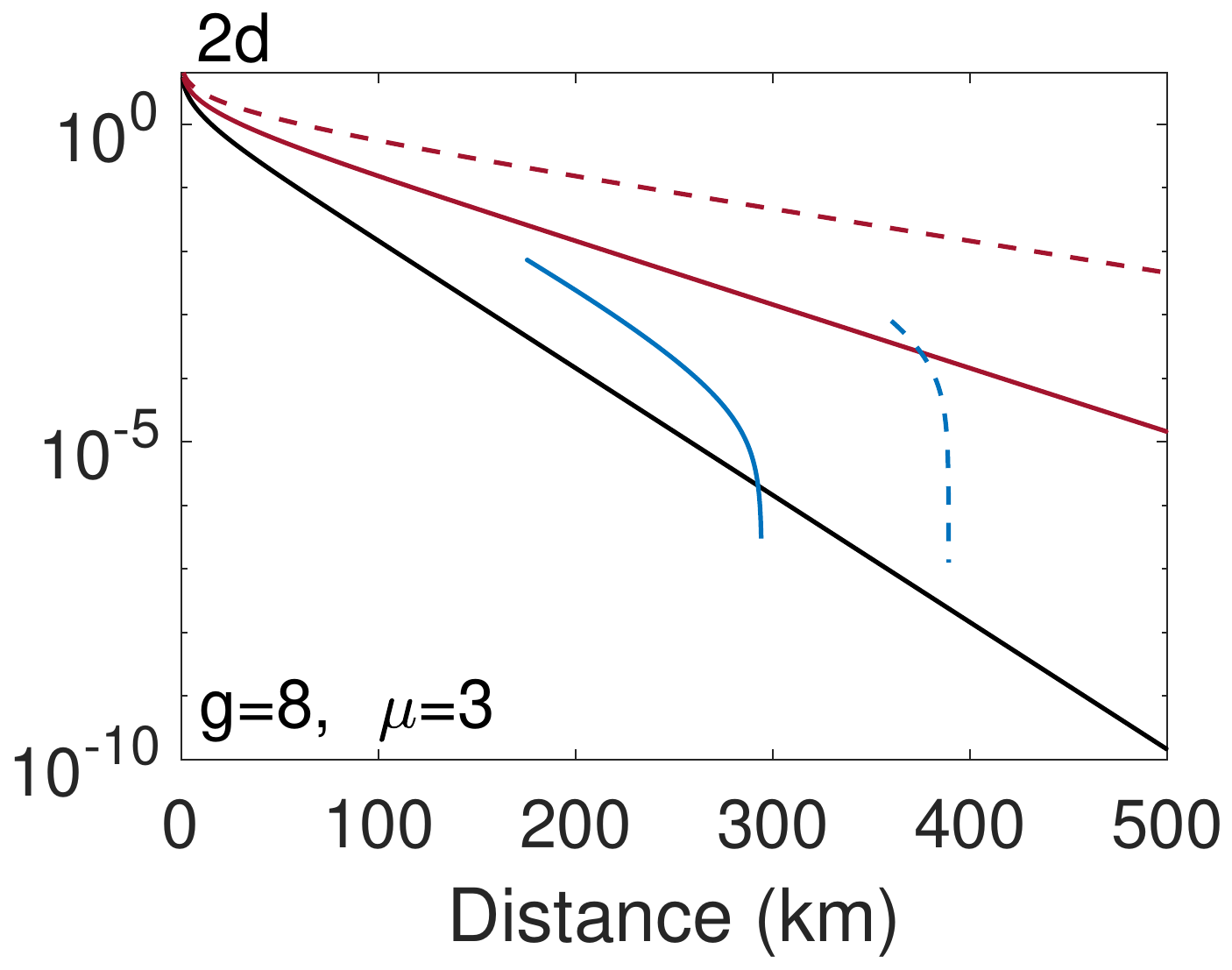}
    \end{subfigure}
       \caption{\textbf{Performance of the quantum repeater with ideal quantum memories.} Reverse coherent information (bits per channel use) versus Alice–Bob’s distance (km) at the loss rate of 0.2~dB/km, and \textbf{zero (row 1)} and \textbf{0.005~snu (row 2)} excess noise. In each subplot, we compare PLOB bound (solid black), RCI for our repeater setup at depth $n=1$ (solid blue) and $n=2$ (dashed blue), and the ultimate end-to-end repeater rates, from Ref.~\cite{Pirandola:EndtoEnd2019}, at depth $n=1$ (solid red) and $n=2$ (dashed red). Subplots (a)-(d) correspond to different amounts of  amplification gain and initial modulation variance. We here assume ideal quantum memories.}
   \label{fig:RCIvsDist}
\end{figure*}

\subsection{Optimal achievable rates for repeater chains} 
Here we represent proper estimates for the secret key rate for the repeater system we described. 
Note that we have effectively reduced the end-to-end repeater problem to a point-to-point one, where the CMs given in Eqs.~\eqref{CM:recursive} and \eqref{CM:QRQM} would suffice to derive suitable bounds. 
To achieve this goal, we use the notions of  coherent information (CI) and reverse coherent information (RCI) of a bosonic channel~\cite{GarciaPatron:RCI2009,Pirandola:RCI2009}. As well, we have all the material to compute the relative entropy of entanglement~\cite{Vedral:REE2002}, but CI and RCI meet the needs of this work. 

Take a bipartite maximally entangled state $\Psi_{\mathbf{AB}}$ of two modes $A$ and $B$. 
The Choi state of a channel, $\mathcal{E}$, is defined by the output state that is obtained after mode $B$ propagates through it: $\rho_{\mathcal{E}}:=\mathcal{I}\otimes \mathcal{E}(\Psi_{\mathbf{AB}})$. CI and RCI of the channel are then, respectively, given by 
\begin{align}
I_{\rm CI}(\mathcal{E})=S[\mathrm{tr}_\mathbf{A}(\rho_\mathcal{E})]-S(\rho_\mathcal{E}),
\end{align}
and
\begin{align}
I_{\rm RCI}(\mathcal{E})=S[\mathrm{tr}_\mathbf{B}(\rho_\mathcal{E})]-S(\rho_\mathcal{E}) ,
\end{align}
where $S(\cdot)$ is the von Neumann entropy.
These two, hence, provide a lower bound on the maximum achievable rate, $R$ \cite{Pirandola_PLOB17}; in fact, we have that
\begin{align}
\max \{I_{\rm CI}(\mathcal{E}), I_{\rm RCI}(\mathcal{E})\} \leq R.
\end{align}  

For bosonic systems, where $\Psi_{\mathbf{AB}}$ is described by the general form of a covariance matrix 
$\mathbf{V}$ in Eq.~\eqref{CM:TMSV}, it can be shown that \cite{Pirandola_PLOB17}
coherent and reverse coherent information are, respectively, given by
\begin{align}
{I}_{\rm CI}(\mathbf{V})=h(a)-h(\nu_-)-h(\nu_+)
\end{align}
and
\begin{align}
{I}_{\rm RCI}(\mathbf{V})=h(b)-h(\nu_-)-h(\nu_+),
\end{align}
where $\nu_{\pm}$ are symplectic eigenvalues of $\mathbf{V}$, 
$h(x)=\frac{x+1}{2}\log_2\frac{x+1}{2}-\frac{x-1}{2}\log_2\frac{x-1}{2}$.
We remark that in all cases considered in this work, RCI stands as the lower bound on the optimal key rate. In addition, if the CM is time dependent---as it is the case when non-ideal quantum memories are in use---so are its eigenvalues. Thus, naturally, entailing Eq.~\eqref{CM:QRQM}, RCI becomes a function of time. 

Let us also point out that when allowing NLAs to operate, the whole repeater system becomes non-deterministic, even-though CV Bell detections are deterministic. 
We set the maximum probability of success for the NLAs, $1/g^2$, as can be achieved via ideal NLAs \cite{Pandey_Qlimits_Amp2013,Blandino:idealNLA2012}.
{This allows us to to obtain an optimistic estimate of the performance of the quantum repeater chain.}
Note that because we let basic links to run in parallel, as previously discussed, the total success probability of the repeater chain is yet $1/g^2$. Not to mention, if basic links are run in a series scenario, we would have had $(1/g^2)^N$ as total success probability and consequently come by a very poor performance.

\section{RESULTS and DISCUSSION}
\label{sec:results}
In this section, we compute reverse coherent information for the quantum repeater setup, whereby we discuss the effect of loss, noise, and non-ideal quantum memories. More importantly, we show that the proposed setup can in fact approach the ultimate rates potentially achievable by means of a repeater chain.  

\subsection{Performance with ideal quantum memories}
Figure~\ref{fig:RCIvsDist} shows RCI versus total repeater distance at few repeater depth $n$. Here, in order to learn the effect of NLAs, loss, and excess noise, we assume ideal quantum memories (we will later study the performance of the repeater with non-ideal quantum memories). 

Figure~\ref{fig:RCIvsDist}(1a) corresponds to the case of \textit{bare} basic links (each comprising a TMSV state and a thermal-loss channel, but without an NLA) connected via CV Bell measurements.
This figure confirms that, in order to outperform fundamental rates, adding other components to the basic link is an ineluctable fact.
In this work, the vital device we add to a basic link is an NLA, the lack of which prevents one to even beat the PLOB bound, let alone the end-to-end repeater rates. 
Now let increase the amplification gain; see Fig.~\ref{fig:RCIvsDist}(1b-1d).
We observe that the quantum memory and NLA-assisted rates outdistance not only no-NLA ($g=1$) rates, but also the PLOB bound, at all repeater depths considered, for some values of amplification gain $g>1$. Also, at longer distances, the repeater rates get parallel to the PLOB bound at long distances.

In fact, the least our repeater setup does is beating the PLOB bound, even when the communications link is noisy; see Fig.~\ref{fig:RCIvsDist}(row 2). We see that excess noise make the key rates drop after some distance; however, by applying a higher amplification gain, we would be able to securely distribute a key at higher distances. To our knowledge, this is the first theoretical proof that the PLOB bound can be outperformed by a chain of CV repeaters in a noisy regime.

{We remark that all plots presented in Fig.~\ref{fig:RCIvsDist} respect the equivalent condition for NLA-assisted CV-QKD links, i.e., $\gamma_g<1$. We elaborate on this point in Sec.~\ref{subsec:CapReaching}. }

In addition, we observe that the higher the repeater depth, the better the performance of the CV repeater setup at long distances. Here, we moreover observe an already known---in the context of point-to-point CV-QKD---feature of NLAs, that is, they are not functioning at short distances \cite{Blandino:idealNLA2012,Ghalaii:JSTQE2020,Ghalaii:JSAC2020}. Indeed, one may prefer not to use them at short distances, which stands by the fact that one would expect quantum repeaters to work successfully at long distances.
As we show in the following, these rates put a benchmark on the obtainable quantum repeater key rates that are potentially achievable with the most convenient joint CV measurement.

\begin{figure}[t]
	\centering
	\includegraphics[scale=0.6]{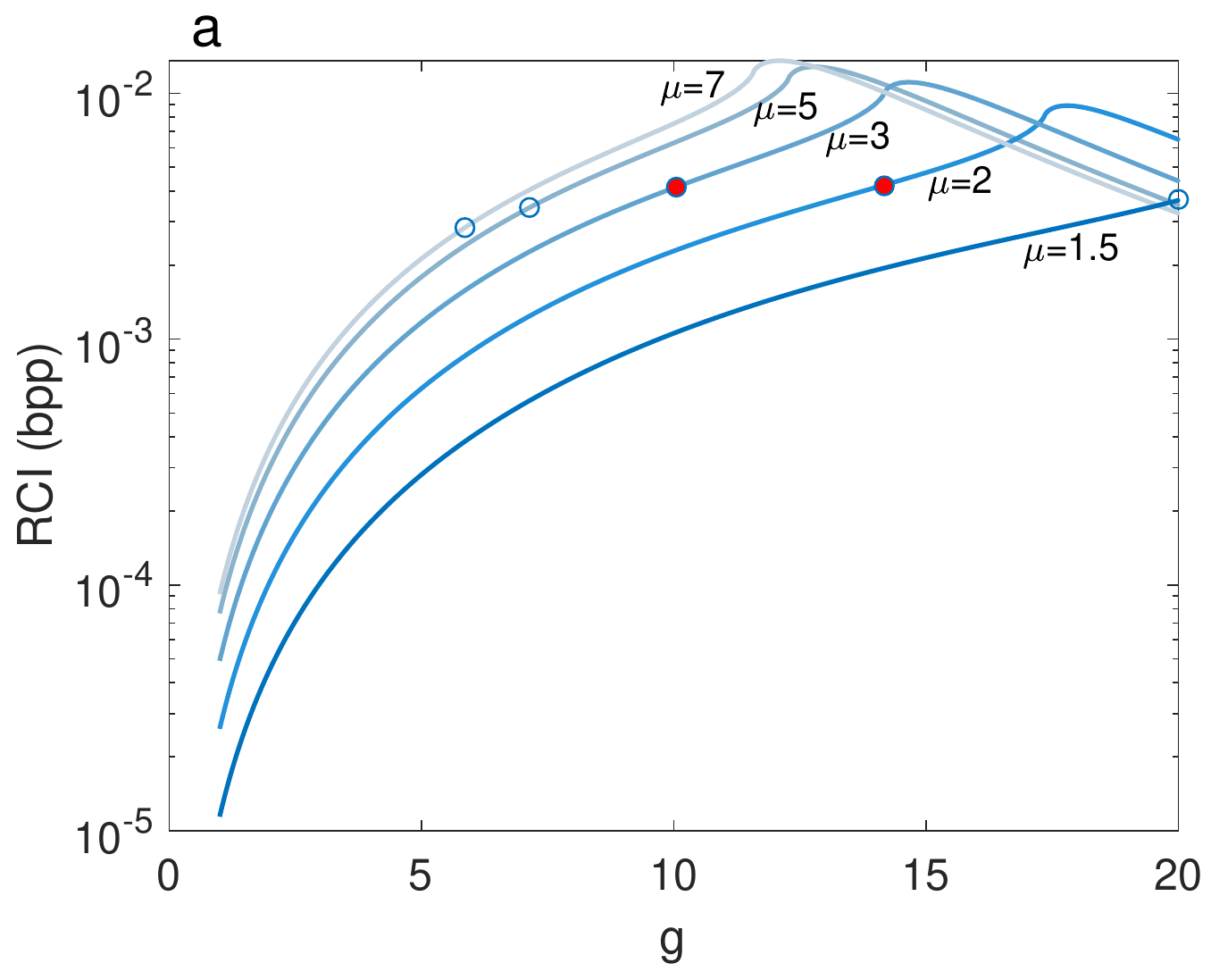} 
	\includegraphics[scale=0.6]{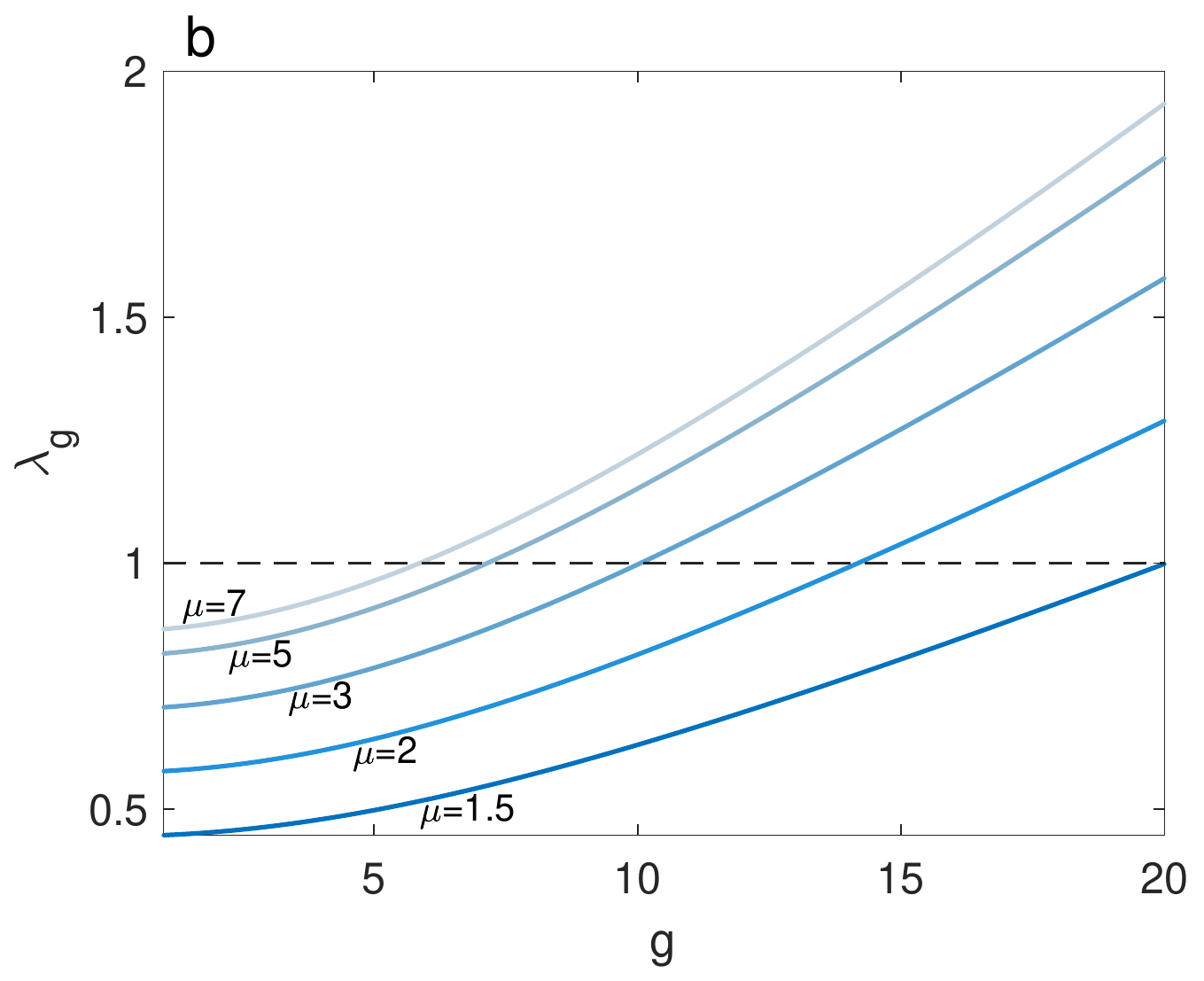}
	\caption{\textbf{(a)} Reverse coherent information and \textbf{(b)} $\lambda_g$ versus amplification gain $g$, at total distance $L=200$~km, zero channel excess noise, and repeater depth $n=1$.
Each curve is tagged with the initial modulation variance $\mu$. The rates after the circles in \textbf{(a)}, which correspond to the $\lambda_g>1$ in \textbf{(b)}, are considered invalid.} 
	\label{fig:optg}
\end{figure}

\subsection{Approaching quantum repeater capacities}
\label{subsec:CapReaching}
As the most interesting result of our study, we show that our repeater proposal can provide us with key rates close to the ultimate end-to-end rates discovered in Ref.~\cite{Pirandola:EndtoEnd2019}.
This can be obtained by optimizing the amplification gain of NLAs and initial modulation variance at each distance. 
As seen in Fig.~\ref{fig:RCIvsDist}, our key rates start closely below the ultimate bounds. At which distance this closeness happens depends on amplification gain $g$ as well as input variance $\mu$. Thus, we could make an optimization over $g$ and $\mu$, at each distance, to see how close can our rates come to the repeater capacities. 

Figure~\ref{fig:optg}(a) shows the lower bounds versus amplification gain $g$, at several values of initial modulation. Here, we have fixed the end-to-end distance $L=200$~km, put zero excess noise, and repeater depth $n=1$, with ideal quantum memories. 
Note that this figure should be taken with some caution since, as previously mentioned, some parts of the generated data may not be reliable \cite{Blandino:idealNLA2012,Ghalaii:JSTQE2020}, i.e., where $\lambda_g>1$.
The point where these reliable distances lays depends on both $g$ and $\mu$, such that higher amplification gains would require lower values of initial modulation variance; this is shown in Fig.~\ref{fig:optg}(b). The fact that we should respect $\lambda_g<1$ implies that the shown rates in Fig.~\ref{fig:optg}(a) are invalid, at a fixed $\mu$, for large $g$'s, indicated by circles on each plot.
It is seen that the maximum key rate, marked by a red-filled circles, can be achieved by either $\mu=3$ and $g=10.05$, or $\mu=2$ and $g=14.17$ (hollow circles are slightly lower that the two red-filled ones). Both configurations would make the optimum rate obtainable that is less than one order of magnitude falls short of the ultimate rate ($1.45\times10^{-2}$) for repeater depth $n=1$.

In above, we assumed a pure-loss channel assisted with ideal NLAs and quantum memories.
We remark that, naturally, with low-quality NLAs and/or low-quality quantum memories, or if the communications link is noisy, the rates would deviate from the potentially obtainable bounds.
In particular, in the following, we explore this deviation caused by non-ideal quantum memories. 

\begin{figure}[t]
	\centering
	\includegraphics[scale=0.6]{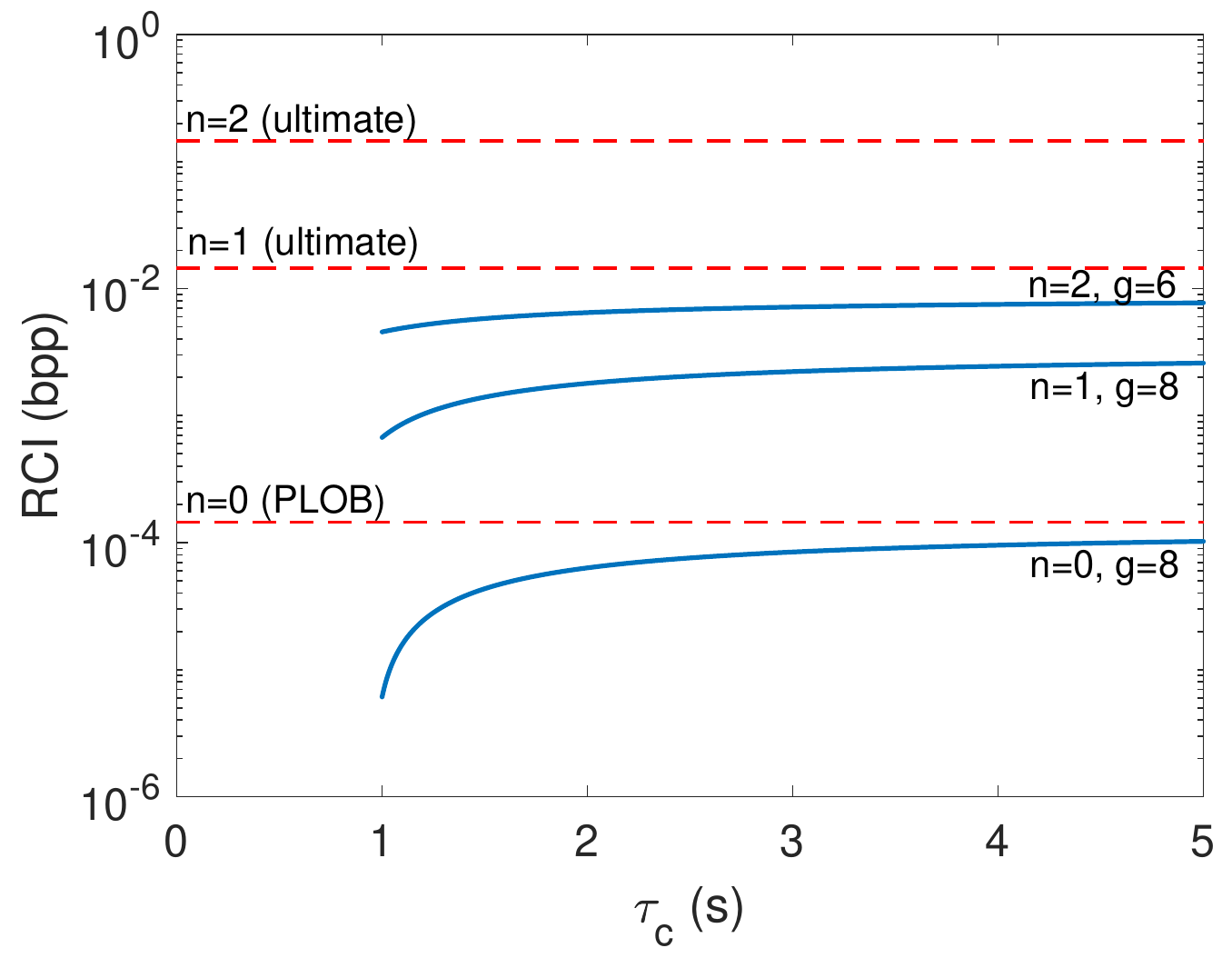}
    \caption{\textbf{Performance of the quantum repeater with non-ideal quantum memories.}
    Reverse coherent information as a function of quantum memories' coherence time, at the loss rate of 0.2~dB/km, 200~km distance, and 0.005~snu channel excess noise. At all cases, we have set excess noise added by quantum memories $\xi_{\rm QM}=0.005$~snu.
    We compare our rates, at different amounts of  amplification gain and repeater depth, with ultimate repeaterless PLOB and end-to-end repeater bounds.}
	\label{fig:RCIvsCohTime}
\end{figure}

\subsection{Performance with non-ideal quantum memories}
We now discuss the effect of realistic CV quantum memories, for which we represented a model, on the lower bounds of our repeater chain. 
For fixed distance $L=200$~km, Fig.~\ref{fig:RCIvsCohTime} shows how the repeater rates compare with ultimate bounds, at few repeater depths, while a noisy and lossy (that is, finite coherence time) quantum memories are in use.
Note that for fixed repeater distance, depth, and amplification gain, the time parameter $t_{\rm h}$ is a constant number.

Figure~\ref{fig:RCIvsCohTime} also makes clear that quantum memories with relatively short coherence times fail to offer a sensible advantage. However, that depends on repeater depth as well as NLA's amplification gain (for each curve we choose the initial variance that gives the highest rate). 
For instance, in this particular case that we examine here, by using a couple of quantum memories with $\tau_{\rm c}>1$~seconds, a repeater chain of depth $n=1$ can achieve rates about one order of magnitude less that the ultimate bound, even-though both the link and memories are noisy. 
We remark that current technology has not yet meet quantum memories that can keep quantum properties for such long times, though a huge effort has been put, both theoretically and experimentally, aiming at the development of quantum memories \cite{Lvovsky:OptQMs2009,Simon:QMReview2010,Jensen:CVQMs2011,Clausen:QMs2011,Saglamyurek:QMs2011}.  

Moreover, for this special case of $L=200$~km, we observe that by going to a higher repeater depth (here $n=2$), one can obtain a slightly higher key, though it is relatively far from its corresponding ultimate rate. Nevertheless, one can inspect from Fig.~\ref{fig:RCIvsDist} that at longer distances, such as 400~km a higher repeater depth may be required. 
In other words, at a fixed channel distance and excess noise, and when quantum memories with certain coherence time are available, we can tell how many repeater stations and how much NLA gain should be used in order to achieve the highest key rate.

\section{CONCLUSION}
We have studied the secret key rates that are achievable by a quantum repeater chain based on NLAs. 
While using typical continuous-variable components, we have proved that our repeater structure is able to surpass the fundamental PLOB limit \cite{Pirandola_PLOB17}, even in a noisy regime of operation. 
Remarkably, we have further shown that the end-to-end repeater-assisted capacities \cite{Pirandola:EndtoEnd2019} can be approached by our repeater structure.
Our rates can outdistance PLOB bound even with non-ideal quantum memories, too, i.e., those that are noisy and have a finite coherence time. To do this feat, and to learn a more realistic account of quantum memories, we proposed a model, which is based on decoherence of bosonic states in a thermal bath \cite{Adesso:QMdecoh2006}.
Our realistic model of a continuous-variable quantum memory behaves like a thermal-loss channel that attenuates the signal, depending on coherence time of the memory, and also adds some excess noise to it. 
Our calculations provide us with the most suitable repeater configuration in terms of depth and amplification gain, once quantum memories with a certain coherence time and excess noise are specified.

{It is worth noting that because of using ideal NLAs our repeater setup is not a totally realistic scenario, though its other parts, i.e., the entanglement source and measurement relay, are readily implementable with current technologies. At the same time, note that the rates in Ref.~\cite{Pirandola:EndtoEnd2019} are obtained by the most arbitrary powerful protocols under general routing strategies, such that it treats a quantum repeater chain very generally and gives no clue towards a physical implementation. Actually, a real-world implementation of repeater chains that can offer repeater obtainable capacities is a substantial problem that, we believe, would need a great effort to be done in future. This current work can be considered as one (first) step towards a fully realistic implementation of such repeater chains where the NLAs are non-ideal, as well. Additionally, it was not trivial that NLAs, either ideal or not, can help to outdistance the PLOB bound and, more importantly, approach repeater capacities. Therefore, by choosing ideal NLAs, we let ourselves to seek the best possible results one can get from such a design. Indeed, one would obtain a poorer performance from non-ideal NLAs. In words, we could consider realistic amplifiers such as quantum scissors, but that would have not end up with what we were seeking for at the first place. Certainly, it is interesting to see how much real-world, non-ideal, NLAs would effect the rates—this, we have left as future research.}

\section*{Acknowledgments}
This work has been sponsored by  the European Union
via \textquotedblleft Continuous Variable Quantum
Communications\textquotedblright\ (CiViQ, Grant agreement No. 820466) and the EPSRC via the UK
Quantum Communications Hub (Grant No. EP/T001011/1).

\section*{Appendices}

\appendix 

\section{Covariance matrix of basic links}
\label{app:triplet}
A CV quantum communications link equipped with an ideal NLA, described by $\mathpzc{E}(\mu,\eta,\xi,g)$ and shown in Fig.~\ref{fig:basiclink_relay}(a), is studied in Ref.~\cite{Blandino:idealNLA2012}, where it is proved that such a system is equivalent to a no-NLA link but now with a modified set of parameters
\begin{align}
\label{equivalent_eqs}
\begin{cases}
\mu_g= \frac{1+\lambda_g^2}{1-\lambda_g^2}, ~~~ \lambda_g= \sqrt{\frac{\mu-1}{\mu+1}\frac{2-\eta (g^2-1)(\xi-2)}{2-\eta(g^2-1)\xi}} \\
\eta_g= \frac{\eta g^2}{1+\eta g^2[\eta(g^2-1)(\xi-2)\xi/4-\xi+1]} \\
\xi_g= \xi- \eta(g^2-1)(\xi-2)\xi/2,
\end{cases}
\end{align}
such that $\mathpzc{E}(\mu_g,\eta_g,\xi_g,1)\equiv \mathpzc{E}(\mu,\eta,\xi,g)$.
Therefore, the corresponding CM elements of the basic link are given by
\begin{align}
\begin{cases}
\mathbf{a}_0=\mu_g \\
\mathbf{b}_0= \eta_g\mu_g + (1-\eta_g) + \xi_g\\
\mathbf{c}_0=\sqrt{\eta_g(\mu_g^2-1)} .
\end{cases}
\end{align}

\section{Post-relay recursive covariance matrix}
\label{app:CVRelay}
Take two separate identical CMs as below
\begin{align}
\label{initialCM}
\mathbf{V}_{\rm a_1b_1}=\mathbf{V}_{\rm a_2b_2}=
\left(\begin{array}{cc}
a \mathbf{I} & c \mathbf{Z} \\
c \mathbf{Z} &  b \mathbf{I}
\end{array}\right).
\end{align}
The compound CM of these two---correspond to a CV repeater with depth $n=1$---reads
\begin{equation}
\mathbf{V}_{\rm a_1b_1a_2b_2}=
\left(\begin{array}{cccc}
a \mathbf{I} & c \mathbf{Z} & 0 & 0 \\
c \mathbf{Z} &  b \mathbf{I} & 0 & 0 \\
0 & 0 & a \mathbf{I} & c \mathbf{Z} \\
0 & 0 & c \mathbf{Z} &  b \mathbf{I} 
\end{array}\right),
\end{equation}
which, after a permutation of modes, can be written as
\begin{equation}
\mathbf{V}_{\rm a_1b_2b_1a_2}=
\left(\begin{array}{cccc}
a \mathbf{I} & 0 & c \mathbf{Z} & 0 \\
0 &  b \mathbf{I} & 0 & c \mathbf{Z} \\
c \mathbf{Z} & 0 & b \mathbf{I} & 0 \\
0 & c \mathbf{Z} & 0 & a \mathbf{I} 
\end{array}\right) \equiv
\left(\begin{array}{ccc}
\mathbf{V}_{\rm a_1b_2} &  \mathbf{C}_1 & \mathbf{C}_2 \\
\mathbf{C}_1^T & \mathbf{B} & \mathbf{D} \\
\mathbf{C}_2^T & \mathbf{D}^T & \mathbf{A} 
\end{array}\right),
\end{equation}
where different blocks are given by
\begin{align}
\mathbf{A}=a\mathbf{I}, ~~
\mathbf{B}=b\mathbf{I}, ~~ 
\mathbf{D}=0,
\end{align}
and
\begin{align}
\mathbf{C}_1=\left(\begin{array}{c}
c\mathbf{Z} \\
0
\end{array}\right), ~~
\mathbf{C}_2=\left(\begin{array}{c}
0 \\
c\mathbf{Z} 
\end{array}\right), ~~
\mathbf{V}_{\rm a_1b_2}=
\left(\begin{array}{cc}
a \mathbf{I} & 0 \\
0 & b \mathbf{I}
\end{array}\right).
\end{align}

As a result of applying the CV Bell detection in Fig.~\ref{fig:basiclink_relay}(b) on modes $\rm b_1$ and $\rm a_2$, the CM of the conditional state is given by the equation \cite{Spedalieri:BellLike2013}
\begin{align}
\mathbf{V}_{\rm a_1b_2|\gamma}=\mathbf{V}_{\rm a_1b_2} - \frac{1}{2\det \mathbf{\Upsilon}} \sum_{j,k=1}^{2} \mathbf{C}_j(\boldsymbol{\upomega}_j^T\mathbf{\Upsilon}\boldsymbol{\upomega}_k)\mathbf{C}_k^T, 
\end{align}
where 
\begin{align}
\mathbf{\Upsilon}=\frac{1}{2}(\mathbf{Z}\mathbf{B}\mathbf{Z}+\mathbf{A}-\mathbf{Z}\mathbf{D}-\mathbf{D}^T\mathbf{Z}) ,
\end{align}
and
\begin{align}
\boldsymbol{\upomega}_1=\left(\begin{array}{cc}
0 & 1 \\
1 & 0
\end{array}\right), ~~~ 
\boldsymbol{\upomega}_2=\left(\begin{array}{cc}
0 & 1 \\
-1 & 0
\end{array}\right).
\end{align}
 
After some algebra, we find the following expression for the post-relay state's CM
\begin{align}
\mathbf{V}_{\rm a_1b_2|\gamma}=
\left(\begin{array}{cc}
\tilde{a} \mathbf{I} & \tilde{c} \mathbf{Z} \\
\tilde{c} \mathbf{Z} &  \tilde{b} \mathbf{I}
\end{array}\right), ~~~
\begin{cases}
\tilde{a}=a - \frac{c^2}{a+b}  \\
\tilde{b}=b - \frac{c^2}{a+b}   \\
\tilde{c}= \frac{c^2}{a+b} .
\end{cases}
\end{align}

Since the resultant CM still has the initial standard from in Eq.~\eqref{initialCM}, one can straightforwardly deduce the recursive expression in Eq.~\eqref{CM:recursive} for higher repeater depths.

\bibliography{biblio}

\begin{thebibliography}{46}
\bibitem [{Op_()}]{Op_times}%
  \BibitemOpen
  \href@noop {} {\bibinfo  {journal} {These include, among others, the time an
  NLA and/or a Bell measurement need to herald successful events, and
  transmission delay $L/c$, where $L$ is the distance between two stations and
  $c$ is the speed of light}\ }\BibitemShut {NoStop}%
\bibitem [{TMS()}]{TMSVdB}%
  \BibitemOpen
  \href@noop {} {\bibinfo  {journal} {For a TMSV state with variance $\mu$, the
  amount of two-mode squeezing in dB is given by
  $-10\log_{10}(\mu-\sqrt{\mu^2-1})$}\ }\BibitemShut {NoStop}%
\end{thebibliography}

\end{document}